\documentclass[11pt,a4paper]{article}
\usepackage[margin=1in]{geometry}
\usepackage{graphicx}
\usepackage{booktabs}
\usepackage{authblk}
\usepackage{hyperref}
\usepackage{xcolor}

\usepackage[textsize=tiny]{todonotes}

\usepackage{changes}
\definechangesauthor[name=huseyin, color=blue]{huseyin}
\definechangesauthor[name=cristiano, color=green]{cristiano}
\definechangesauthor[name=valentina, color=purple]{valentina}
\definechangesauthor[name=celian, color=red]{celian}

\title{X-DigCheck: Co-Evolving Application Profiles and Knowledge Graphs, Demonstrated on the RTI Documentation of Rupe Magna}

\author[1]{Hüseyin Erdoğan\thanks{Email: \texttt{huseyin.erdogan@unibo.it}, ORCID: 0000-0002-2965-0918.}}
\author[2]{Valentina Presutti\thanks{Email: \texttt{valentina.presutti@unibo.it}, ORCID: 0000-0002-9380-5160.}}
\author[1]{Cristiano Putzolu\thanks{Email: \texttt{cristiano.putzolu@unibo.it}, ORCID: 0000-0002-6910-6474.}}
\author[3]{Célian Ringwald\thanks{Corresponding/lead author. Email: \texttt{celian.ringwald@unibo.it}, ORCID: 0000-0002-7302-9037. Authors are otherwise listed alphabetically by family name.}}

\affil[1]{Department of History and Cultures, Alma Mater Studiorum -- University of Bologna, Bologna, Italy}
\affil[2]{Department of Modern Languages, Literatures and Cultures, Alma Mater Studiorum -- University of Bologna, Bologna, Italy}
\affil[3]{Department of Computer Science and Engineering (DISI), Alma Mater Studiorum -- University of Bologna, Bologna, Italy}

\date{}

\begin{document}

\maketitle

\begin{center}
\fbox{\parbox{0.92\linewidth}{\small
\textbf{Author preprint.} This is the authors' own preprint version of a
paper accepted to the ISWC 2026 Companion Volume, October 25--29, 2026,
Bari, Italy. Copyright for this paper belongs to its authors. Use permitted
under Creative Commons License Attribution 4.0 International (CC~BY~4.0).}}
\end{center}

\begin{abstract}
We demonstrate \textbf{X-DigCheck}, a domain-independent environment for
building and maintaining application profiles as they co-evolve with the data
they describe. Profiles developed against a fixed ontology quickly drift from
the schema they were meant to capture. X-DigCheck treats profile
construction as a continuous ontology--data \emph{co-evolution} loop: data are
lifted into RDF against the profile, checked through competency questions and
SHACL, and the resulting reports jointly drive revisions of the ontology,
mappings, constraints, and graph. The loop is agnostic to the domain and to the pipeline
that produces the graph. We validate and demonstrate the tool in the cultural
heritage domain, on the construction of \textbf{RupeMagna-RTI}, the first
Reflectance Transformation Imaging (RTI) specialisation of the Cultural
Heritage Survey ODP (CHS-ODP), aligned with CIDOC-CRM/CRMdig, ArCo, CHAD-KG,
and Getty~AAT, with \textbf{semRTI} as the lifting pipeline of this use case.
The demonstration lets visitors run one full turn of the loop --- on the
shipped Rupe~Magna (Grosio, Italy) RTI survey, or on a profile and graph of
their own --- executing the competency-question and SHACL checks live and
reading the bidirectional coverage report that flags modelling gaps and stale
assumptions. The result is a portable co-evolution
environment for profile engineering, together with a reusable RTI application
profile produced through it. A screencast of the demonstration is available at
\url{https://zenodo.org/records/22210609}.
\end{abstract}


\bigskip

%%%%%%%%%%%%%%%%%%%%%%%%%%%%%%%%%%%%%%%%%%%%%%%%%%%%%%%%%%%%%%
\section{Introduction}
Archaeological documentation increasingly combines heterogeneous acquisition
techniques, including laser scanning, photogrammetry,
and Reflectance Transformation Imaging (RTI)~\cite{malzbender2001ptm}, a
non-invasive computational photography technique that captures an object from
a fixed viewpoint under varying light directions, enabling interactive
relighting that reveals fine surface details such as engravings. Each
technique introduces specific parameters, workflows, and representations. A
single monolithic ontology cannot easily balance domain expressiveness and
interoperability. Application profiles provide a more flexible approach: rather than
extending a domain ontology directly, a profile specialises it for a
particular community of practice --- selecting the terms, cardinalities,
and value spaces relevant to a specific use case, which can then be
validated via a shapes layer (e.g.\ SHACL) against instance data.
We use \emph{ontology}/\emph{vocabulary} for this
domain-level conceptual layer, \emph{application profile} for its
community-specific specialisation, and \emph{shapes} for the validation
layer that checks instance data against it; X-DigCheck operates on the
latter two, treating the profile as the object under revision and the
shapes as its (partly derived, partly authored) validation surface. We build
on shared conceptual foundations, specialising the Cultural Heritage
Survey Ontology Design Pattern (CHS-ODP)~\cite{cappa2026chsodp} as the shared conceptual foundation for such profiles.
 
However, application profiles evolve with both domain knowledge and produced
data. Treating them as static extensions risks misalignment between modelling
assumptions, workflows, and validation requirements. We therefore frame
profile construction as an ontology--data \emph{co-evolution} process, where
profiles, mappings, constraints, and knowledge graph instances evolve through
iterative modelling, generation, validation, and revision cycles.

The contribution of this paper is \textbf{X-DigCheck}\footnote{\url{https://codeberg.org/infinity-project/x-digcheck}}, an environment that
makes this loop operable: it assesses a profile against the knowledge graph
generated from it through competency questions, SHACL validation, and
bidirectional coverage analysis, and turns the resulting reports into
actionable revisions of ontology, mappings, constraints, or data. The loop is
independent of the application domain and of the pipeline producing the
graph: any workflow generating RDF against an evolving profile can feed it.
We validate and demonstrate X-DigCheck on the engineering of
\textbf{RupeMagna-RTI}, the first RTI specialisation of CHS-ODP, with
\textbf{semRTI} --- a declarative lifting pipeline for RTI acquisition
outputs --- instantiating the generation side of the loop for this use case.
Rather than reporting the loop as a finished result, the demonstration lets
an attendee run one turn of it and read the drift live.
% \added[id=celian]{Every mechanism described below is available both through
% the browser portal and headlessly through a companion CLI client, so the
% same checks can be scripted, repeated, and fed into downstream tooling
% (Section~\ref{sec:workflow}).}

\paragraph{Related work.}
CIDOC~CRM~\cite{doerr2003cidoc} and ArCo~\cite{carriero2019arco} provide general-purpose cultural heritage models, but they partially address acquisition-specific workflows. CHS-ODP provides the reusable conceptual core specialised in the validation use case, while CHAD-KG~\cite{BARZAGHI2026} contributes the digitisation module integrated into the resulting knowledge graph. In the broader context of Cultural Heritage data, the Europeana Data Model (EDM)~\cite{EDM24} also plays an important role by providing a framework and a set of profiles for the interoperability and aggregation of cultural heritage resources. Building on this ecosystem, the use case presented here yields a first prototype of an RTI profile supporting the representation and validation of the RTI archaeological survey of Rupe~Magna~\cite{erdogan_2025_21300400}.

Approaches such as Astrea~\cite{cimmino2020astrea} and OWL2SHACL~\cite{owl2shacl} derive SHACL constraints from OWL ontologies--- automating the profile-to-shapes step in isolation, while ABSTAT~\cite{palmonari2015abstat} and OOPS!~\cite{povedavillalon2014oops} focus on ontology quality assessment and dataset profiling at the ontology layer. X-DigCheck differs from these single-layer approaches by focusing on the continuous alignment across the profile, shapes, and data layers at once --- between an application profile and the generated knowledge graph, making this alignment observable as it evolves across successive regenerations. Building on iterative ontology engineering methodologies such as SAMOD~\cite{peroni2016samod} and eXtreme Design~\cite{blomqvist2010xd}, and extending the previously proposed TestALod interface~\cite{carriero2019arco}, which focused exclusively on ontology validation through competency questions (CQs), we introduce a user-oriented, that could also support CLI automation, interface supporting the validation and evolution of profiles, mappings, constraints, and generated data. This approach complements recent continuous integration solutions such as OLIVAW~\cite{robert2025olivaw}, an implementation of ACIMOV methodology~\cite{hannou2023acimov}, by extending ontology evolution practices towards the co-evolution of the complete knowledge graph generation pipeline.

%%%%%%%%%%%%%%%%%%%%%%%%%%%%%%%%%%%%%%%%%%%%%%%%%%%%%%%%%%%%%%
\begin{figure}[h!]
\centering
\includegraphics[width=\linewidth]{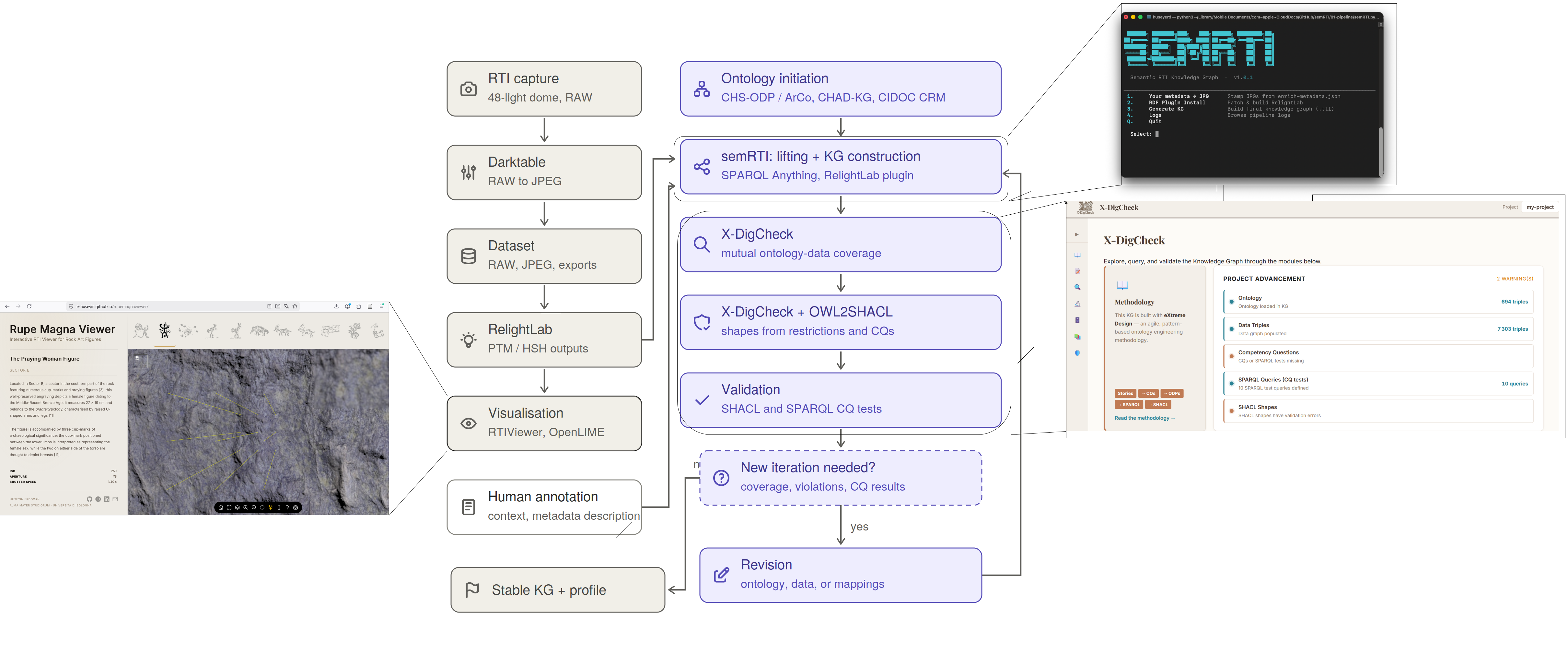}
\caption{Overview of the co-evolution workflow. Acquisition outputs and manual field annotations are lifted into RDF by \textbf{semRTI} according to the ontology profile. The resulting knowledge graph is evaluated with \textbf{X-DigCheck}, where ontology-knowledge graph alignment, competency questions, and SHACL validation (generated through OWL2SHACL and complemented with manually authored shapes) produce reports that guide revisions of the ontology profile, mappings, or data before regenerating the knowledge graph.}
\label{fig:workflow}
\end{figure}
\section{The Co-Evolution Workflow}
\label{sec:workflow}
We consider knowledge graph construction as a co-evolution workflow in which changes to the ontology profile, lifting rules, validation requirements, and source annotations are progressively integrated and evaluated. On the generation side, \textbf{semRTI}, based on SPARQL~Anything, lifts RTI acquisition outputs into RDF against the ontology profile in the demonstrated use case; any other pipeline producing RDF against an evolving profile --- whether based RML, or custom code, and whatever the domain of the data --- can take semRTI's place in the loop. On the validation side, \textbf{X-DigCheck} provides the domain-independent environment at the centre of this loop, assessing the resulting graph against the profile and turning the outcome into actionable revisions. Figure~\ref{fig:workflow} provides an overview of this iterative process.

\paragraph{semRTI: acquisition and semantic lifting in the use case.}

In the demonstrated use case, survey data are acquired under an LED hemispherical dome, developed from RAW images, and processed with RelightLab~\cite{ponchio2019relight,ponchio2024relightlab} to generate RTI files. \textbf{semRTI} is a Python pipeline that transforms these RTI outputs into a FAIR-compliant RDF/Turtle knowledge graph. The pipeline combines two complementary inputs. The first consists of the RTI products together with their source images and associated metadata. The second is a manually curated annotation block that captures contextual information unavailable through automatic extraction, such as the physical object under study, its position on the rock surface, acquisition conditions, and field observations.
Metadata embedded in EXIF headers and RelightLab JSON descriptors are queried \emph{in situ} using SPARQL~Anything~\cite{asprino2023sparqlanything}. RDF is generated directly through CONSTRUCT mappings written against the ontology profile, making semantic construction an integral part of the lifting process rather than a subsequent transformation step. In addition, a RelightLab plugin produces Turtle sidecar files describing each PTM/HSH rendering, including processing parameters and provenance links to the original acquisition. The generated graph is finally processed through OWL reasoning to materialise the entailments prescribed by the ontology profile. Since every asserted triple remains traceable to its originating EXIF field, JSON attribute, or manual annotation, the entire knowledge graph can be regenerated automatically whenever the profile or mappings evolve. Once semRTI has produced or updated this graph, \textbf{X-DigCheck} takes over as the validation and revision environment described next.

\paragraph{X-DigCheck: validation and co-evolution tracking.}  The application is a self-contained Docker environment built around an Apache Jena Fuseki SPARQL~1.1 triple store, deployable either as a full browser portal or headlessly through a CLI client that mirrors every browser action for scripted, repeated, or CI-driven runs. Fuseki ships as the bundled default store, that could be used to load KG data or a sample of it, and every validation mechanism (coverage analysis, competency questions, SHACL) talks to the graph exclusively through the standard SPARQL~1.1 protocol; pointing X-DigCheck at an external endpoint works with any SPARQL~1.1-compliant sparql endpoint, not only with the local Fuseki KG. Beyond knowledge graph exploration
through YASGUI, the iterative
co-evolution of an application profile and the generated knowledge graph
proceeds through three validation mechanisms. First, X-DigCheck analyses
profile--knowledge graph alignment by detecting unused ontology elements,
insufficiently described data, and modelling mismatches. These findings
guide profile extensions and refinements of the lifting mappings. Second,
it supports competency question execution following eXtreme Design
(XD)~\cite{blomqvist2010xd}, an agile, pattern-based ontology engineering
methodology in which competency questions are elicited before any axiom is
written and compiled directly into SPARQL regression tests.
SPARQL queries
encoding expected profile capabilities can be run against successive
knowledge graph versions and recorded in validation reports, providing
regression testing for profile and mapping changes. Third, X-DigCheck
integrates SHACL validation using shapes automatically derived from logical
axioms through OWL2SHACL and manually defined shapes for constraints beyond
OWL expressivity. The resulting reports identify inconsistencies and
missing information, providing feedback for refining the profile, mappings,
and source data, and can be exported as structured Markdown for downstream
consumption --- e.g.\ as input to an external coding agent tasked with
fixing the flagged elements. None of these mechanisms depends on how the
instance data is stored: the profile and shapes are managed as local files,
while the underlying knowledge graph data can be loaded locally or, for an
evolving environment, queried live from any external SPARQL~1.1 endpoint
--- so the same three checks apply unchanged to any profile--graph pair.
Together, they establish the feedback loop enabling profile and knowledge
graph co-evolution. Furthermore, this process can be used on a project developed using Xtrem Design's iterative method, by importing a Git project.
%%%%%%%%%%%%%%%%%%%%%%%%%%%%%%%%%%%%%%%%%%%%%%%%%%%%%%%%%%%%%%
\section{The Validation Use Case: RupeMagna-RTI and Its Knowledge Graph}
\label{sec:profile}
\paragraph{Demo video.}
A three-minute screencast of the walkthrough described below is available
at \url{https://zenodo.org/records/22210609}, showing the live competency-question and SHACL run and the
coverage report on the shipped Rupe~Magna graph.
\paragraph{Use case and dataset.}
The demonstration ships the artefacts produced by ten turns of the loop
above\footnote{\url{https://cringwald.codeberg.page/RupeMagnaOnto/dataviz/dashboard.html}}
on the Rupe~Magna (Grosio, Italy) RTI survey initiated
in~\cite{erdogan_2025_21300400}, which documents Alpine rock art dating from
the Late Neolithic to the Iron Age. The September~2025 campaign produced 40
RTI sessions covering 11 engraved figures, each captured through 48 images
under a custom 48-LED dome. Twenty-four of these datasets are published,
together with their processing and publication layers, under the persistent
namespace \url{https://w3id.org/rupemagna-rti/ontology#}; a full regeneration
of the graph currently yields roughly 148k triples. This use case exercises
every mechanism of the loop --- lifting from heterogeneous sources, derived
and manual shapes, coverage analysis across a modular profile. It thereby
serves as the validation context for X-DigCheck, and the profile it produced
is a reusable result in its own right.

\paragraph{Application Profile structure.}
The profile evolved from a monolithic model into six domain modules\footnote{\url{https://cringwald.codeberg.page/RupeMagnaOnto/ontology/}} built on a
shared core: \emph{survey}, \emph{agent--role}, \emph{equipment and raw
capture}, \emph{observation and photographic documentation},
\emph{digitisation process}, and \emph{DCAT publication}. Modules import only
the core and remain independently complete, and are synchronised through
automated checks. These checks revealed modelling issues not visible through
data coverage alone. Digitisation is represented as a provenance chain (survey,
RAW development, RTI fitting, export), aligned with CIDOC~CRM/CRMdig, ArCo,
CHAD-KG, DCAT, OntoPiA, and Getty~AAT. Measurements are stratified into three
layers: \emph{exposure and equipment} at session level, \emph{per-image}
observations of file properties, and \emph{processing} outputs covering RTI
and web parameters. Each figure is modelled as a chain of derived versions
from raw captures to relightable visualisations, linked to the software
executions that produced them and typed as both \texttt{dcat:Distribution}
and \texttt{chs:Result}, with explicit licensing for published renderings.

\begin{figure}[h!]
\centering
\includegraphics[width=0.7\linewidth]{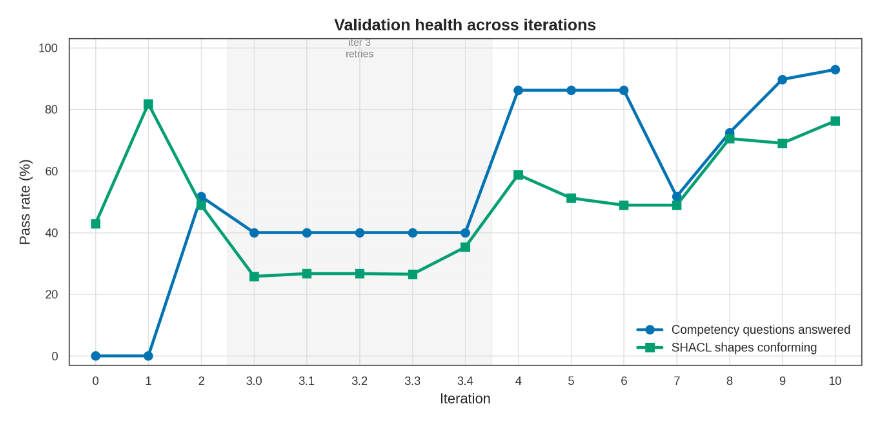}
\caption{X-DigCheck validation loops of the competency-question and SHACL
results}
\label{fig:validation}
\end{figure}

\paragraph{What the loop produced.}
The eight competency questions inherited from CHS-ODP were expanded into
twenty-nine profile-specific ones covering seven themes: survey and method,
observation semantics, agent roles, equipment and exposure, measurement
layers, derivation lineage, and publication structure. Early versions encoded
protocol assumptions in OWL restrictions (e.g., one equipment configuration
per survey, or 48 observations per collection); the modular rewrite moved all
cardinality out of the ontology into SHACL, as contextual validation rather
than ontological commitment. Two shape sets are therefore maintained in
parallel: one derived by OWL2SHACL from the profile's own axioms, one authored
manually for Rupe~Magna-specific requirements. This lets the profile define
meanings and relations without redefining imported terms or embedding
instance data, while keeping project-specific conformity checks explicit. At
the current stable iteration (Figure~\ref{fig:validation}), 26 of the 28 executable competency questions
pass (93\%) and 32 of 42 shapes conform (76\%), with 8\% of profile classes
and 3\% of properties never instantiated and no undeclared terms remaining.
Several engineering steps were LLM-assisted --- drafting shapes, writing and
repairing mappings, updating queries --- each admitted only after passing the
competency-question and SHACL test bank described above.

%%%%%%%%%%%%%%%%%%%%%%%%%%%%%%%%%%%%%%%%%%%%%%%%%%%%%%%%%%%%%%
\section{Demonstration}
%\label{sec:demo}
\paragraph{Setup and walkthrough.} semRTI and the RupeMagnaViewer\footnote{\url{https://e-huseyin.github.io/rupemagnaviewer/}} are run to show the data and their transformation into RDF; alongside them, X-DigCheck is run to show the process leading to the production of the knowledge graph and the adaptation of CHS-ODP into a specialised RTI application profile.

The demonstration proceeds in four steps. First, attendees load the
git-tracked RupeMagnaOnto project into X-DigCheck and inspect the
statistics produced across its full agile development cycle. Second, they focus on a specific iteration and
execute competency questions and SHACL validation, comparing profile-derived constraints against contextual, manually authored shapes that capture application-specific requirements beyond OWL expressivity, and export the resulting validation report. Third, attendees switch to a sandbox use case, restricted to the alignment between a knowledge graph and a profile supplied independently of RupeMagna-RTI --- either from another use case or from their own data --- showing that the same checks generalise beyond this paper's dataset. Fourth, we show that every validation available through the UI can equally be run headlessly through the \texttt{xdigcheck\_cli.py} client: inspecting ontology coverage and model drift --- identifying unused profile terms and data elements not represented in the profile --- is as
scriptable as it is browsable. These discrepancies indicate whether the next
refinement should target the ontology, mappings, constraints, or source
data.
\paragraph{Resources.}
All software used in the demonstration is available as
open-source prototypes: semRTI and its RelightLab RDF
plugin\footnote{\url{https://codeberg.org/erdoganhuseyin/semRTI}}, and
X-DigCheck\footnote{\url{https://codeberg.org/cringwald/X-DigCheck}}, with
the X-DigCheck project template also available\footnote{\url{https://codeberg.org/cringwald/X-DigCheck_template}} for reuse on other use cases.
 All data produced for the demonstration is
openly available under CC~BY~4.0 and open-source licences: the competency
questions, the RupeMagna-RTI application profile, the shapes, and the
resulting RDF dataset\footnote{\url{https://codeberg.org/cringwald/RupeMagnaOnto}}.

%%%%%%%%%%%%%%%%%%%%%%%%%%%%%%%%%%%%%%%%%%%%%%%%%%%%%%%%%%%%%%
\section{Discussion and Conclusion}

The co-evolution loop confirms that ontology engineering and knowledge graph
construction evolve together. Iterative validation exposed both transformation
artefacts (incorrect mappings, missing properties, absent inverses) and
field-data gaps, allowing the two to be distinguished and resolved through
successive iterations. The temporary coverage regression following the
digitisation-module rewrite reflected the expected lag between an enriched
TBox and its ABox rather than a modelling failure. Conformance recovered
once the graph was regenerated.

This work introduces X-DigCheck, an environment for iterative ontology--data
co-evolution, validated and demonstrated through the engineering of RupeMagna-RTI,
the first RTI application profile derived from CHS-ODP, using semRTI as the
lifting pipeline for the use case. Rather than treating validation as a final
verification step, the proposed approach integrates competency questions,
SHACL validation, and ontology--knowledge graph coverage analysis throughout
profile development. The validation context is cultural heritage, and in
particular RTI survey documentation; the mechanisms themselves are
domain-agnostic, and the environment is portable to any setting where a
knowledge graph is generated against an evolving profile: biomedical or
bibliographic data pipelines could equally feed the same loop unchanged. The results
are a practical, reusable methodology and tool for maintaining alignment
between ontologies and their knowledge graphs as they evolve, and, as a
by-product of the validation, a well-documented and reusable semantic profile
for RTI surveys.

While demonstrated on a single RTI use case, this work establishes a
foundation for deriving further reusable, domain-specific application
profiles from shared conceptual cores such as CHS-ODP, across acquisition
techniques and beyond cultural heritage. Future work on the use-case side
focuses on extending the RTI profile to support acquisition quality
assessment, archaeological interpretation, and links to existing
archaeological literature. X-DigCheck is moving towards an ontology
development environment that tracks coverage, SHACL validation, and
competency-question results across Git revisions --- in the spirit of ACIMOV~\cite{hannou2023acimov} and OLIVAW~\cite{robert2025olivaw} --- while
remaining independent of the mapping technology (e.g., SPARQL~Anything or RML). However, we are also considering incorporating the mapping files into future work so that we can, in particular, include any potential errors caused by this "facelift" step in the generated reports. A longer-term objective is to combine the co-evolution loop with agentic LLM assistance for mapping repair, shape generation, and query maintenance, while preserving executable validation criteria for every generated artifact.

\section*{Acknowledgments}
This work was carried out within the INFINITY project, funded by the
European Union's Horizon Europe research and innovation programme under
grant agreement No.~101233051.
We thank the anonymous reviewers for their valuable feedback, which encouraged us to extend X-DigCheck with a CLI mode and to integrate the full development cycle starting from a Git project.

\section*{Declaration of use of Generative AI}
Generative AI tools were used for language editing and, as reported in
Section~\ref{sec:profile}, within the engineering process itself: drafting
SHACL shapes from the profile's constraints, writing and repairing the
SPARQL~Anything CONSTRUCT mappings, and updating the competency-question
queries. All such output was reviewed by the authors and admitted only
after passing the validation and competency-question test bank described
above. The conceptual contributions, engineering decisions, validation
methodology and conclusions are entirely the authors'.

\bibliographystyle{plain}
\bibliography{references}

\end{document}